\documentclass[11pt,preprint]{aastex}

\shorttitle{Asphericity of SN 2002ic}
\shortauthors{Chugai N.N. \& Chevalier R.A.}

\begin{document}

\title{MODERATE ASPHERICITY OF THE SN 2002ic CIRCUMSTELLAR ENVELOPE}

\author{Nikolai N. Chugai}
\affil{Institute of Astronomy, RAS, Pyatnitskaya 48, 109017 Moscow, Russia}
\email{nchugai@inasan.ru}

\and

\author{Roger A. Chevalier}
\affil{Department of Astronomy, University of Virginia, P.O. Box 400325,
Charlottesville, VA 22904-4325, USA}
\email{rac5x@virginia.edu}

\begin{abstract}

The polarization of SN~2002ic interacting with a dense circumstellar 
envelope is calculated in the context of 
the asymmetric version of a previously proposed
spherical interaction model. 
The circumstellar envelope is taken to be oblate.
The observed polarization \citep{Wa04} can be reproduced 
for an aspect ratio of 0.65-0.7 assuming inclination angles $>60^{\circ}$.
This model predicts a weak sensitivity of the line profiles to 
the orientation, in agreement 
with the absence of significant variations of the line profiles
among SN~2002ic-like supernovae.  We propose a test for distinguishing 
between the binary and single star progenitor scenarios based 
upon the polarization distribution function for the growing sample of 
these events. 

\end{abstract}

\keywords{stars: mass-loss --- supernovae: general --- supernovae:
individual (\objectname{SN 2002ic})}

\section{INTRODUCTION} \label{sec-intro}

A new variety of Type Ia supernova (SN) embedded 
in a dense circumstellar (CS) envelope has been recognized 
\citep{Ham03}. The generic example is SN~2002ic. 
Another example is the recently discovered SN~2005gj \citep{Pri05,Ald06}.
Two other suspected members of this new family are 
SN 1997cy \citep{Tur00} and SN 1999E \citep{Rig03}. 
Although both were discovered long after the 
explosion ($t>100$ d), their close relation to SN~2002ic according to
spectra, line profiles, and light curves \citep[cf.][]{Den04}
is clear.  The mass of the 
CS envelope derived from the luminosity powered by the interaction  
during $\sim 500$ days is several $M_{\odot}$ 
\citep[][hereafter CCL04]{Den04,Chu04a}.

The origin of these events is a challenging problem because 
existing evolutionary scenarios for SNe~Ia do not predict
such dense CS envelopes. 
Although single degenerate binary with supergiant secondary 
 or single star (SN~1.5) progenitor scenarios
are conceivable \citep{Ham03}, both face 
the serious problem of synchronization between the phase of the rapid
loss of a massive hydrogen envelope and the explosion \citep{LR04, CYu04}. 
A binary scenario proposed by \citet{HP06} suggests that a main 
sequence secondary experiences a Roche lobe overflow at the time 
when the white dwarf companion has attained the Chandrasekhar limit. 

An understanding of the properties of the CS envelope 
might elucidate the issue. Specifically, a strong deviation from 
spherical symmetry would support 
the binary over the single star
scenario. The detection of the intrinsic polarization 
of SN~2002ic at the level of $p\approx0.8$\% \citep{Wa04} 
indicates asymmetry. Wang et al.  conjectured that the CS medium 
around this supernova is strongly aspherical and makes up a dense 
equatorial disk with an opening angle of the order of a few degrees.
Such structure, if confirmed, would have 
important implications for the evolutionary scenario.
However, the spherical model is consistent with the observed
line profiles and quasi-continuum of SN~2002ic (CCL04).
Moreover, all three supernovae with available 
spectra (SN~2002ic, SN 1997cy, SN 1999E) 
show similar shapes and widths of spectral lines at similar 
epochs, which would not be expected in the case of a strongly 
aspherical interaction (CCL04). 
 
These issues show the need 
to develop a model for the polarization of SN~2002ic.
This paper aims to answer the questions:
(1) what is the 
minimal degree of asphericity implied by the polarization data, and 
(2) is the minimal asphericity consistent with the spherical 
appearance suggested by the spectroscopic arguments?
Our study is based on the spherical model of SN~2002ic 
proposed earlier for the light curve and spectrum 
of SN~2002ic (CCL04), modified here by introducing 
asphericity of the CS envelope. The polarization is calculated 
for a number of aspherical cases using the Monte Carlo technique.
The results of polarization modelling are discussed along with 
the modelling of H$\alpha$ profiles in the context of  
comparison with available observational constraints.

\section{MODELLING OF POLARIZATION}

\subsection{Model}

We share the view of \cite{Wa04} that the polarization 
of SN~2002ic is caused by Thomson scattering of photons 
in an aspherical CS envelope. A possible contribution of 
the asphericity of the initial SN ejecta is neglected, given the fact 
that the polarization of most SNe~Ia is weak ($<0.2-0.3$\%), 
 except for possibly the outermost layers \citep{Wa03}
which contain a dynamically unimportant mass of the ejecta. 

Given the difficulty of self-consistent 
modelling of the hydrodynamics and the optical light 
for the case of the interaction of a SN~Ia with an aspherical CS envelope,
we follow a simplified approach. 
We rely on the 
model of SN~2002ic interaction with a spherical CS envelope 
computed in the thin shell approximation (CCL04). 
The model has a flat CS density distribution
at $r<3\times10^{15}$ cm, a power law drop
$\rho\propto r^{-1.4}$ for $3\times10^{15}<r<2.6\times10^{16}$ cm, and a 
steep decay $\rho\propto r^{-10}$ at $r>2.6\times10^{16}$ cm.
The total mass of the CS envelope in the spherical model is $1.6~M_{\odot}$.
The radius of the thin shell at the epoch of the polarization 
observations ($\sim 244$ d) is $1.9\times10^{16}$ cm.
We modify the spherical model by assuming 
that the CS envelope is aspherical and the isodensity surfaces are 
oblate spheroids with an aspect ratio 
$b/a<1$ (where $a$ and $b$ are the major and minor semi-axes).
The interaction zone, including the forward shock, reverse shock and 
cool dense shell (CDS) acquires, on the contrary, the shape 
of a prolate spheroid with the long axis along the polar axis (Fig. 1). 
Below we omit a thin layer of hot gas in 
the reverse shock that contributes neither to the optical 
radiation nor to the Thomson scattering. 

The CDS formed by the shocked ejecta is responsible for the optical 
radiation except for the H$\alpha$ line (CCL04).
The layer between the CDS of radius  $R_{\rm c}$
and the forward shock of radius  $R_{\rm s}$
($R_{\rm c}<r<R_{\rm s}$) contains the shocked CS gas, which 
is assumed to be homogeneously distributed with radius, but with the 
dependence on the polar angle $\theta$ obtained from the column 
density of the swept-up CS gas. 
The CS density distribution along the minor axis 
is assumed to be the same as in the spherical model. Given the spheroidal 
shape of isodensity surfaces, the total mass of the CS envelope 
exceeds that of the spherical model by a factor $(a/b)^2$. 
CCL04 argued that the CDS may be fragmented and 
mixed with the forward shock gas. We considered two extreme 
cases: an unmixed case in which the CDS is unmixed and 
all the photons are emitted at the radius $R_{\rm c}$, and a mixed case 
in which the emitting CDS material is homogeneously mixed in the 
forward shock region, $R_{\rm c}<r<R_{\rm s}$, and did not find 
a noticeable difference in polarization between these cases.  
Here we present results only for the unmixed case. 

The shape of the CDS is approximated by 
a prolate spheroid with the angular dependence of the 
radius, $r=R_{\rm c,p}(1-\zeta\sin^2\theta)$, where 
the asphericity parameter is related to the polar and equatorial 
radii of the CDS, $\zeta=1-R_{\rm c,e}/R_{\rm c,p}$.
To estimate the polar radius 
$R_{\rm c,p}$ and velocity $v_{\rm p}$ of the CDS along with 
the corresponding 
equatorial values, $R_{\rm c,e}$, $v_{\rm e}$, we used the spherical 
dynamical model (CCL04) computed for the CS medium along the polar and equatorial 
density distributions, respectively, assuming  
radial flow. The radial assumption is reasonable for moderate values
of the asymmetry \citep{BLC96}.
We found the shape and velocity asphericity parameters 
to be $\zeta=0.09$ and $\xi=1-v_{\rm e}/v_{\rm p}=0.14$ 
for $b/a=0.7$ on day 244.
The case $b/a=0.65$ gives $\zeta=0.11$ and $\xi=0.16$. 

The emissivity depends more strongly on the polar angle than
the shape and velocity.
The ratio of the equatorial to polar bolometric fluxes 
for $b/a=0.7$ found from the thin shell approximation 
model is $1+\eta=1.66$, i.e. $\eta=0.66$.  For $b/a=0.65$,  
the model gives $\eta=0.78$.
The expansion velocity of the CDS is larger at the pole, 
while the X-ray emissivity is larger at the equator where the 
column emission measure is larger. The variation of the 
X-ray flux with polar angle can be approximated by the expression 
$F\propto (1+\eta\sin^2\theta)$. The same expression is used 
for the emissivity. 

The photons emitted by the CDS experience Thomson scattering with electrons 
in the (i) ionized unshocked CS gas with a H abundance of $X=0.7$;
(ii) shocked CS gas in the forward shock; and (iii)  unshocked SN ejecta.
The luminosity of the narrow H$\alpha$ emission line 
on day 272 (assuming an explosion on JD 2452585) according to \cite{KM04} 
is $\sim5\times10^{39}$ erg s$^{-1}$. A simple estimate based on the 
recombination mechanism shows that the CS gas 
must be fully ionized to reproduce the required luminosity.
The ionization is presumably 
maintained by the absorption of the hard X-ray emission from the 
forward shock with a luminosity 
of $\sim5\times10^{42}$ erg s$^{-1}$ and temperature of $40$ keV (CCL04).
Given the uncertainty of the emission measure 
related to the possible clumpiness we consider two cases:  complete 
ionization (ionization fraction $x=1$) and partial ionization ($x=0.5$). 

In addition to the hot gas, the postshock gas in the forward shock presumably also contains 
a cool ($T\sim10^4$ K) component responsible for the broad H$\alpha$ 
emission line (CCL04).   Pressure equilibrium with the ambient postshock gas
for the typical average preshock density of the CS gas of
$\sim 10^7$ cm$^{-3}$ and a forward shock velocity of 
$\sim 6000$ km s$^{-1}$ (CCL04) implies a density 
for the cool gas of $\sim10^{12}$ cm$^{-3}$ at the relevant epoch. 
Assuming that half the swept-up CS gas ($\sim0.5~M_{\odot}$) 
is in the form of the dense shocked cool gas we find that 
the recombination H$\alpha$ luminosity at the time in question,
$\sim5\times10^{40}$ erg s$^{-1}$ \citep[cf.][]{KM04} 
requires an ionization fraction $x\sim 0.02$.
We approximately take into account the presence of the  H$\alpha$-emitting 
gas in the forward shock by assuming 
that half the swept-up CS gas in the forward shock is neutral.

The determination of the ionization of the unshocked SN ejecta requires a more 
complicated model. We calculated 
the ionization balance in the SN ejecta in the X-ray radiation field 
from the reverse 
shock predicted by the spherical model (CCL04),
assuming that all the absorbed energy is spent on ionization. 
The ejecta are assumed to consist of pure iron. Calculations 
with radiative and dielectronic recombination rates taken from 
\cite{SS82} result in almost the same ejecta ionization
on day 244 for the electron temperatures $T_e=10^5$ K and $10^6$ K.
The ionization stage increases monotonically from 
Fe$^{+}$ in the center to Fe$^{+6}$ in the outer layer of ejecta. The 
uncertainty of the ionization model does not markedly affect  the 
resulting polarization. In the extreme case of  complete ionization
of the ejecta, the polarization is larger by only a factor of 1.3 compared 
to the preferred case of the ejecta ionization computed for $T_e=10^5$ K.
A typical value of the Thomson optical depth for the SN ejecta at the 
relevant epoch is $\tau_{\rm T}\sim0.01$, while for the CS envelope together 
with the forward shock $\tau_{\rm T}\sim0.1$.

\subsection{Results}

The Monte Carlo technique is used to treat the 
transfer of polarized radiation emitted by the CDS. The algorithm 
is similar to that
used by \cite{Angel69}. Two vectors are tracked: 
the electric field and the photon wave vector. 
The direction of the scattered photon is chosen according to the 
dipole law of scattering. Stokes components 
$Q$ and $U$ of the escaping photons 
are summed with the corresponding array elements for a 
certain inclination angle $\psi$.
Light-travel time effects are neglected, which does not introduce a 
significant error.  
The code was tested using available numerical and semi-analytical 
results, including the classical Chandrasekhar-Sobolev solution for 
a Thomson scattering atmosphere. 

The calculated polarization of SN~2002ic on day 244 
as a function of $\cos\,\psi$ 
for the aspect ratios $b/a=0.7$ and 0.65  assuming complete 
ionization ($x=1$) of the CS envelope is shown in Fig. 2
(upper panel). The escaping radiation is polarized along the small axis.
The lower panel shows the effect of lower 
ionization ($x=0.5$ vs. $x=1$) for $b/a=0.65$.
The plot also gives  the reported average polarization 
of SN~2002ic with the r.m.s. deviation \citep{Wa04}. 
For complete ionization 
both aspect ratios are consistent within errors with  
the detected polarization in a range of inclination 
angles $\psi>60^{\circ}$; the case of $b/a=0.7$ demonstrates 
the minimum asphericity compatible with the observations.
In the case of partial ionization ($x=0.5$)
the required asphericity should be somewhat larger, 
$b/a\ge 0.65$.
For $b/a=0.7$ the mass of the CS envelope is
$3.2~M_{\odot}$, i.e., a factor of two 
greater than found in the spherical case (CCL04). 

Using the spheroidal model of the CS envelope and CDS radius according to 
the spherical model (CCL04) we computed the 
polarization at the earlier epochs (100, 150, and 200 days) 
for the case $b/a=0.7$ and $x=1$ in the same way as 
for $t=244$ d. The early polarization is larger (Fig. 3) owing to
more efficient scattering in the envelope with larger optical depth. 
Such behavior is characteristic of the optically thin regime. 

To summarize, the observed polarization of SN~2002ic can be reproduced 
in a model of a spheroidal CS envelope with the density predicted by 
the spherical interaction model (CCL04); the minimum required 
aspect ratio is moderate, $b/a\approx 0.65-0.7$.

\section{ASPHERICITY AND THE H$\alpha$ PROFILE}

We now address the issue of whether the minimum asphericity found from 
the polarization modelling is consistent with the assumption that 
it is common for SN~2002ic-like supernovae and with the fact that  
the known supernovae (2000ic, 1997cy and 1999E) do not 
reveal significant variations of the spectral line profile width and shape
for both H$\alpha$ and Ca\,II 8600 \AA\ at an age of 
$\sim240$ d \citep[e.g.,][]{Wa04}. 

We concentrate on the strong and well delineated H$\alpha$ line.
With spectral resolution of $\sim220$ km s$^{-1}$ the 
H$\alpha$ profile on day 240 is concave-shaped \citep[cf.][]{Wa04},
or, more specifically, the full width at half maximum 
${\rm FWHM}\approx 1600$ km s$^{-1}$ is significantly smaller than the
blue (or red) width at zero intensity (BWZI or RWZI), which is  
$5000-6000$ km s$^{-1}$. This type of profile is 
in apparent contrast with that of other CS interaction supernovae, 
e.g., SN~1987F with a parabolic line H$\alpha$ 
\citep{Fil89, Chu91} or SN~1993J with a boxy H$\alpha$ line
\citep{FMB94} -- in both cases the FWHM exceeds the BWZI.
The high resolution spectrum of SN~2002ic reveals on top of H$\alpha$
a narrow P Cygni line originating from the CS envelope with 
a velocity of $\sim 100$ km s$^{-1}$ \citep{KM04}. 
Here we do not consider the CS component; it is composed of 
at least two velocity components: narrow emission with FWHM=78 km s$^{-1}$ 
and absorption with the blue wing velocity of $\sim250$ km s$^{-1}$.
The interpretation of the CS H$\alpha$ profile is not obvious and 
requires a separate study. 

We follow the conjecture that broad H$\alpha$ is emitted by shocked 
CS clouds in the forward shock (CCL04). The qualitative model suggests 
that CS clouds are crushed by slow radiative cloud shocks; the
shocked clouds are then dispersed by the Kelvin-Helmholtz instability
into a multitude of fragments that are accelerated by the fast 
forward shock flow. This sequence is similar to 
that found in two-dimensional modelling of blast wave 
interaction with 
an interstellar cloud (Klein et al. 1994). As a result of the
interaction of the forward shock with a cloudy CS envelope, a broad 
velocity spectrum ($dM/dv$) of the line-emitting dense cool gas 
presumably forms in the forward shock \citep{Chu04b}.
The minimum velocity corresponds to undisturbed CS clouds, while 
the highest velocity $v_0$ corresponds to the velocity of fragments 
accelerated up to the
velocity of the swept-up shell $v_0\sim 6000$ km s$^{-1}$.
We adopt here a simple parametrization of the 
velocity spectrum of the line-emitting gas 
$dM/dv\propto v$ for $0<v<v_m$ and 
$dM/dv\propto (v_0-v)^q$ for $v_m<v<v_0$, with the maximum at 
$v=v_m$ that presumably corresponds to the typical velocity of
cloud shocks. The free parameters are $v_m$ and $q$.
In the case of a spheroidal CS 
envelope, $v_0$ depends on the polar angle 
$v_0=v_{\rm p}(1-\xi\sin^2\theta)$, where $v_{\rm p}$ is the 
CDS velocity at the pole.
Assuming that the line emissivity per unit mass is constant, 
the resulting profile $F(u)$ (where $u$ is the radial velocity) 
is calculated by integration over the transverse velocity $w$ 
using the relation $v^2=u^2+w^2$:
\begin{equation}
F(u)\propto\frac{dM}{du}=\int\frac{dM}{dv}2\pi wdw\,.
\end{equation}

We restrict ourselves to a qualitative description of the broad H$\alpha$. 
For the moderately aspherical model consistent with the
polarization data ($b/a=0.65$,  $\eta=0.78$, $\xi=0.16$, 
$v_{\rm p}=6000$ km s$^{-1}$) we find that 
the H$\alpha$ FWHM and profile shape in SN~2002ic on day 244 
\citep{Wa04} are basically reproduced if $v_m=0.07v_0$ and $q=0.7$.
The model profiles in this case are shown in Fig. 4 (upper panel)
for the inclinations $\psi$  equal to $0^{\circ}$, $60^{\circ}$ 
and $90^{\circ}$. The profiles for 
$\psi=90^{\circ}$ and $60^{\circ}$ are almost identical, while 
in the pole-on case ($\psi=0^{\circ}$) 
the FWHM is a factor of 1.33 smaller than for $\psi=90^{\circ}$. 
Interestingly, although the velocity is largest at the poles, the profile 
is broader when the SN is viewed in the equatorial plane. 
This effect is due to the higher emissivity at the equator.
The implication is that for a spheroidal CS envelope in 
a sample of SN~2002ic-like supernovae,
the polarization should correlate with the H$\alpha$ width.

To assess whether the variation of FWHM values with the inclination 
in the SN~2002ic model for the moderate asphericity is compatible 
with the range of H$\alpha$ width among three
 SN~2002ic-like supernovae, we note that the maximum FWHM is
shown by SN~1999E with FWHM$\sim1900\pm100$ km s$^{-1}$ 
\citep{Rig03}, while the lowest width H$\alpha$ is in SN~2002ic with 
FWHM$\sim1550$ km s$^{-1}$ \citep{KM04}. The 
ratio of widths is $\sim1.23\pm0.07$, comparable to the 
maximum ratio 1.33 predicted by the model of spheroidal CS envelope 
with $b/a=0.65$. Given only three known events, one may 
conclude in a preliminary way that the hypothesis 
of CS envelopes with moderate asphericity ($b/a\geq0.65$) is  
consistent with both the small variation of the H$\alpha$ 
line profiles and the polarization revealed by SN~2002ic.

Unfortunately, our interaction model is not applicable to 
strongly flattened CS envelopes, so we are not able 
to explore confidently the H$\alpha$ profile behavior in 
such cases. Just to get an idea about the trend, we consider
a CS envelope with an aspect ratio $b/a=0.3$.
For this case the derived asphericity parameters for the 
velocity and emissivity are $\xi=0.616$ and $\eta=0.98$ 
respectively. 
Assuming that the line-emitting gas has the same velocity distribution as 
in the model with $b/a\geq0.65$, we obtain the line profiles for the same 
inclinations shown in Fig. 4 (lower panel). The resulting 
profiles differ in two respects compared to those for $b/a=0.65$: 
(1) they have lower FWHM by a factor of $\sim1.25$; (2) their 
broad wings for large inclinations get weaker and completely 
disappear for $\psi=90^{\circ}$. 
The discovery of SN~2002ic-like supernovae with a ``typical'' luminosity 
but lower than ``normal'' FWHM of H$\alpha$ and strongly 
suppressed broad wings at an age of $\sim 200-250$ days would thus 
indicate   interaction with a strongly flattened CS envelope and a line of 
sight close to the equatorial plane.  Such an event should be accompanied 
by an unusually large intrinsic polarization. Alternatively, 
the absence of such events in a sufficiently large sample of 
SN~2002ic-like supernovae would be evidence 
against the strong asymmetry of the CS envelopes.

\section{CONCLUSION}

The primary goal of this paper was to find the 
minimum asphericity of the CS envelope around SN~2002ic 
compatible with the polarization data. It is demonstrated that 
a spheroidal CS envelope with a moderate aspect ratio 
$0.65-0.7$ and  density and length scale 
parameters of a previous spherical model (CCL04) 
is consistent with the observed polarization for a wide 
range of inclination angles $\psi>60^{\circ}$. We also found 
that if all  SN~2002ic-like supernovae are similar and 
have  CS envelopes with a similar asymmetry ($b/a\approx0.65-0.7$), 
then the expected variation of the H$\alpha$ width due to random 
inclination is moderate and does not contradict the small variation
of the observed H$\alpha$ widths. 
This resolves the apparent disparity between the asphericity of the 
CS envelope implied by the polarization \citep{Wa04} and the 
high degree of symmetry implied by the similarity of line profiles for 
SN~2002ic-like supernovae (CCL04). Yet we stress that our analysis 
and existing observational data do not completely 
rule out the possibility that the CS envelopes 
around SN~2002ic-like events are significantly flattened.

The conclusion that the asphericity of SN~2002ic-like supernovae 
may be modest does not resolve, however, another important issue: 
whether  asphericity of the CS envelope is a generic
property of the CS envelopes around these supernovae. 
There are at least two possibilities:
(1) A single star ($\sim6-8~M_{\odot}$) scenario holds and 
the asphericity of the CS envelope around SN~2002ic is a 
phenomenon perhaps related to the presence of a second 
star in a wide binary system \citep[e.g.,][]{MM99}.
The possibility that 
a single star might retain enough angular momentum in the 
envelope to produce significant asymmetry of the outflow
is also conceivable but seems to be unlikely
\citep[cf.][]{Sok96}. (2) A close binary scenario holds and the 
asphericity is a generic property of SN~2002ic-like supernovae. 

The issue could be resolved through polarization 
observations of SN~2002ic-like supernovae.
The single star option predicts that the probability 
distribution function of polarization ($dN/dp$) 
of SN~2002ic-like supernovae peaks at $p=0$ with a 
weak tail up to $p\sim 1$\% populated by rare cases of 
wide binaries. For the second option, the distribution function
is determined by random inclinations of aspherical CS envelopes 
with a uniform distribution of $\mu=\cos\psi$: 
$dN/dp=(dN/d\mu)|d\mu/dp|$. 
The $p(\mu)$ dependence in Fig. 2 
is approximately described by $p=p_0(1-\mu^s)$, where $s\approx2$. Therefore, 
given $dN/d\mu=const$, the expected distribution in the binary scenario,  
$dN/dp\propto (1-p/p_0)^{-(s-1)/s}$, should be peaked at the 
maximum polarization with a value $\sim1$\%. 
The weak dependence of the polarization on the age 
in the range of $100<t<250$ d (Fig. 3)
implies that the uncertainty of the phase of the polarization observation
by $\pm50$ days is not critical for determining 
the polarization distribution function $dN/dp$. 

The polarization data for a growing number of these 
supernovae thus may become a useful test for both scenarios.
In this regard, polarization measurements of the recent SN~2005gj, 
as well as other newly discovered SN~2002ic-like supernovae,
at an age $100-250$ days are highly desirable.

\acknowledgements
We are grateful to the referee for constructive comments on
the manuscript.
The research was partially supported by RFBR grant 02-17255 (NNC)
and NSF grant AST-0307366 (RAC).

{}

\clearpage

\begin{figure}
\plotone{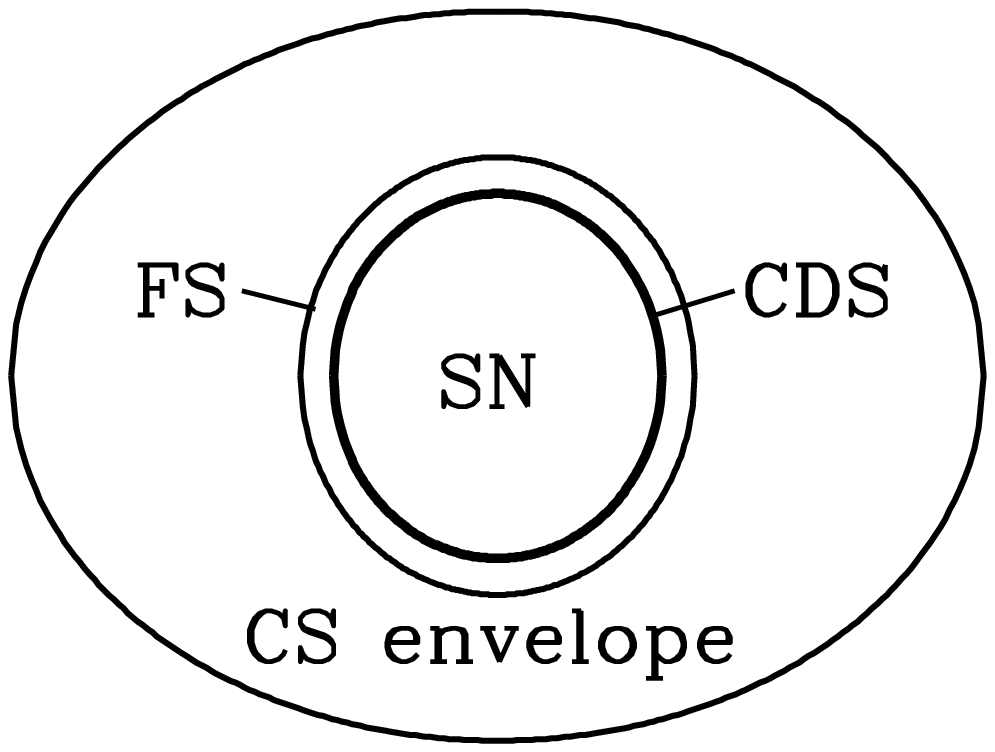}
\caption{Sketch of a supernova in a spheroidal 
  circumstellar envelope. The major components are SN ejecta 
  bounded by a cool dense shell (CDS) and the forward shock (FS) 
  propagating in the spheroidal circumstellar (CS) envelope.}
  \label{f-cartoon}
  \end{figure}

\clearpage

\begin{figure}
\plotone{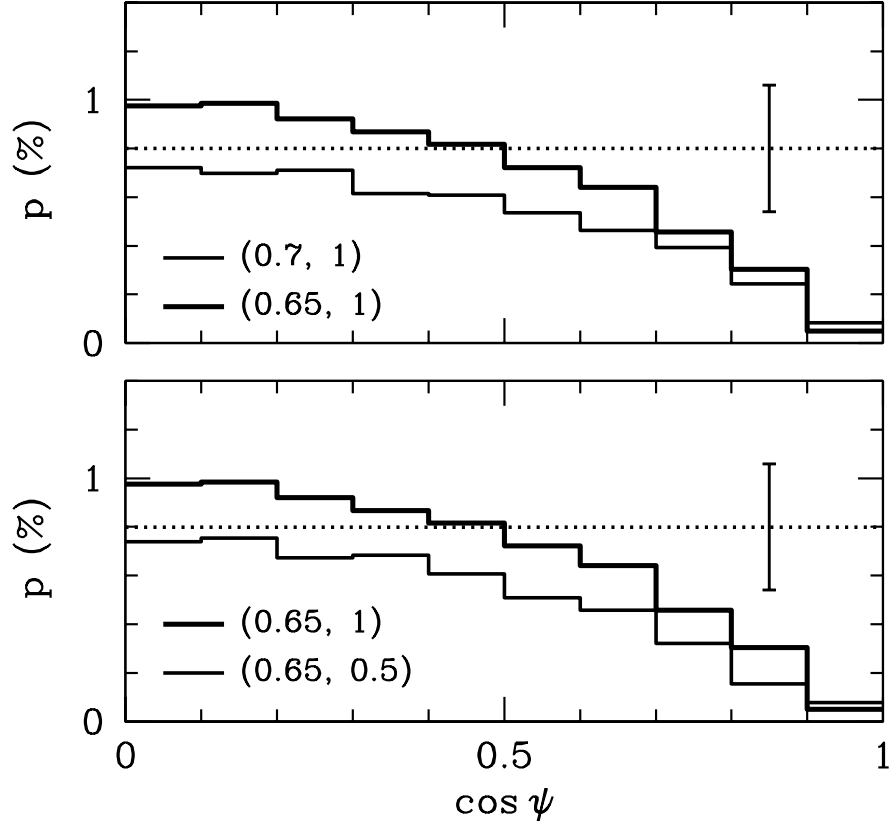}
\caption{Model polarization as a function of cosine of inclination
  angle.  The quantities in parentheses 
  are the aspect ratio $b/a$, and the ionization fraction of the 
  CS envelope $x$.
  The {\em upper} panel shows models for $x=1$ and 
  two values of the aspect ratio.  The {\em dotted} line 
  is the average value of observed polarization (Wang et al. 2004)
  with the bars showing the error. The {\em lower} panel 
  shows the effect of ionization of the CS envelope for $b/a=0.65$.}
  \label{f-polar1}
  \end{figure}

\clearpage

\begin{figure}
\plotone{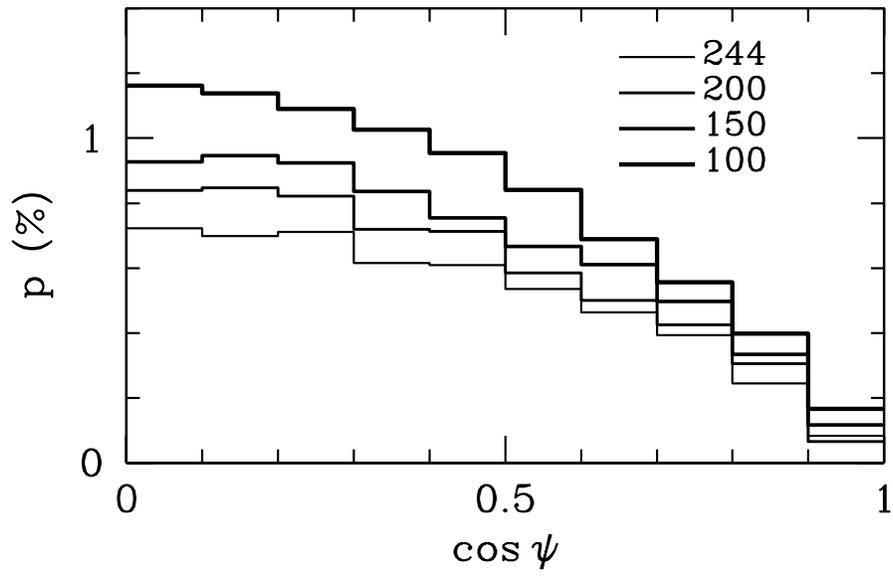}
  \caption{Model polarization between days 100 and 244
  as a function of the inclination for $b/a=0.7$.}
   \label{f-polar2}
  \end{figure}

\clearpage
\begin{figure}
\plotone{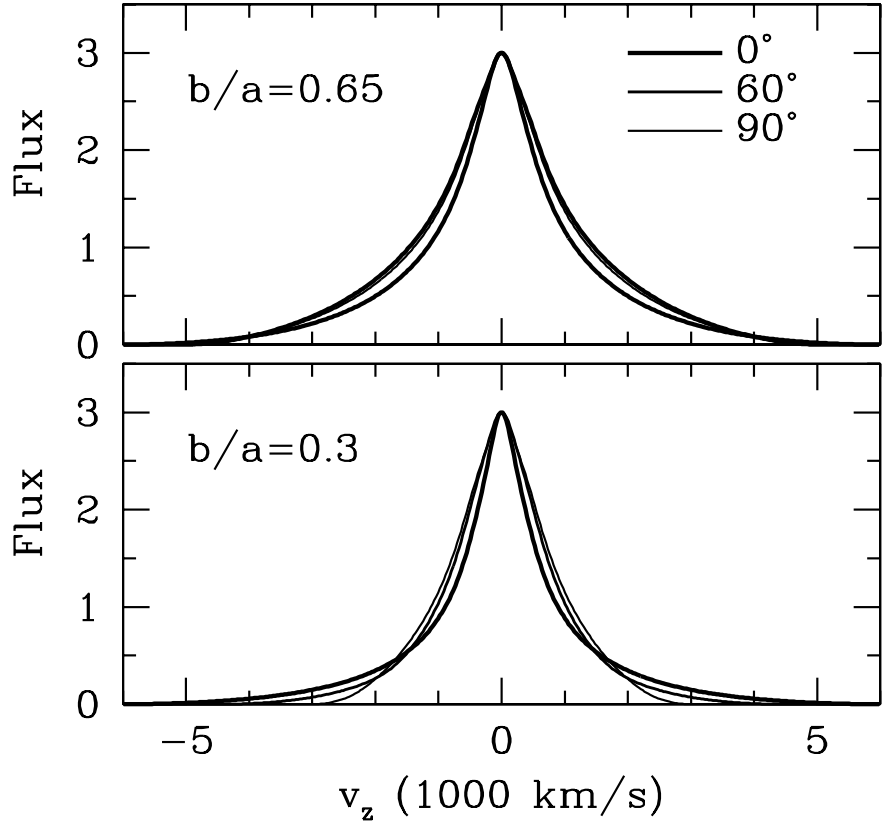}
  \caption{The model of the H$\alpha$ line profile for the 
  aspherical line-emitting shell with two extreme inclinations.
  {\em Upper panel} shows the profile for the case of a spheroidal 
  CS envelope for $b/a=0.65$ (note that the $60^{\circ}$ and $90^{\circ}$ 
  cases are overlapping), while the {\em lower panel} is the same 
  but for $b/a=0.3$.}
  \label{f-profile}
  \end{figure}

\end{document}